# Long-ranged attraction between disordered heterogeneous surfaces

Gilad Silbert[1], Dan Ben-Yaakov[2], Yael Dror[1], Susan Perkin[3], Nir Kampf[1], Jacob Klein[1*]

*1. Department of Materials and Interfaces, Weizmann Institute of Science, Rehovot, 76100, Israel.*

*2. Department of Physics and Astronomy, Tel Aviv University, Ramat Aviv, Tel Aviv 69978, Israel*

*3. Department of Chemistry, University College London, 20 Gordon Street, London, WC1H 0AJ, UK.*

*\*Jacob.klein@weizmann.ac.il*

**The stability of colloidal dispersions, as recognized since the work of Faraday and Graham some 150 years ago, hinges on the balance between attractive and repulsive surface forces(*1*). In aqueous media the electrostatic double-layer repulsion between like-charged surfaces plays the dominant role in the stabilization, keeping particles from aggregating under van der Waals attraction(*1-3*). Such forces have been studied for decades(*3-6*): but while nearly all understanding of the interactions relate to uniformly-charged surfaces, most real surfaces are in fact heterogeneous and disordered. Here we demonstrate that two surfaces covered with random charge patches experience a long-ranged attraction across water that is orders of magnitude stronger than van der Waals forces, and which persists, remarkably, even in the complete absence of any correlations between opposing positive and negative domains (so-called 'quenched disorder'). The origin of this attraction is, as we show, in the counter-intuitive observation that two oppositely-charged surfaces attract each other across water (or other ion-containing liquids) much more strongly than equally-charged surfaces repel, for identical surface separations and charge densities. This striking asymmetry may result in strong, long-ranged attraction between randomly charged surfaces even when they are overall-neutral. It sheds new light on interactions of hydrocolloids and in biological systems, where such surfaces are ubiquitous. It may also account for the long-ranged attraction frequently observed between surfaces coated with surfactants (and with other species) which was long attributed to a hydrophobic surface interaction.**

Heterogeneous surfaces with random positive and negative charge domains were created via an initially-uniform self-assembled surfactant layer which re-arranged with time under water (Methods). Fig. 1 shows typical AFM micrographs of the monolayers, as-freshly-deposited and following immersion in water. As observed previously with many other surfactants(*7-11*), the initially smooth monolayer breaks up with immersion time in water into a random mosaic of positively-charged bilayers (mean thickness 3.2±0.4 nm) on a background of negatively-charged bare mica.

Earlier studies revealed that such surfaces – bearing positively- and negatively-charged domains of random size and distribution – experience an attraction at separations of several tens of nanometers(*7-11*) which is orders of magnitude stronger than can be due to van der Waals forces. This strong, long-ranged attraction was attributed to the correlation of charge domains (positive facing negative) developing on approach of the surfaces(*7-13*). Here we carried out experiments to examine directly the crucial point whether such charge correlations were occurring: Normal forces $F_n(D)$ between the two surfaces at closest distance D apart were determined in a surface force balance (SFB, Methods) in two different modes, illustrated in the insets A and B to fig. 2 . The first, inset A, termed the straight-approach mode, is the one used in all earlier studies of surface forces: The surface separation D is progressively decreased by moving them normally with respect to each other, to construct the force vs. separation profile, $F_n(D)$. In the second mode, the shear-approach, the surfaces are again made to approach, but now the sectored piezoelectric tube on which the upper surface is mounted is used, *during this normal approach*, to apply a rapid lateral (shear) motion parallel to the lower surface (inset B to fig. 2), over a range of frequencies ν and amplitudes *A*. The amplitude, *A*, of this lateral motion (120 – 600 nm) is much larger than the characteristic size of the charge domains (from ca. 10 - 200 nm, fig. 1). The lateral velocity $v_s = 2\nu A$ ranges to values (5.1 μm/sec) that are up to three orders of magnitude larger than the estimated value of $v_{domain}$, the velocity of the charge domains during their assumed motion to correlation, at the onset of attraction(*10*). Such applied lateral motion $v_s \gg v_{domain}$ must thus frustrate any directed translation of the charge domains on either surface to a correlated configuration with the opposing charge regions. If indeed therefore it was the charge correlations postulated earlier that were leading to the long-ranged attraction in the straight-approach mode, then the attractive normal interactions measured using the shear-approach mode should be much weaker, and, in particular, they would onset at a much shorter range. Likewise, as the surfaces come into adhesive contact, correlation between the opposing charges – as on the straight approach - would lead to larger adhesion energies than with the shear-approach where such correlation is absent.

Several typical surface interaction profiles using these two modes are shown in fig. 2, for STAB-coated surfaces that had been immersed water for extended periods. As clearly seen (and suggested also by an earlier preliminary measurement(*14*)), there is no systematic difference within the experimental scatter at a given contact point between the straight approach mode (as used in all earlier studies), and the shear-approach mode designed to frustrate any formation of a correlated (positive-faces-negative) charge distribution: Both show the characteristic long-ranged attraction reported earlier(*7-11*). Likewise, as shown in fig. 3, there is no systematic difference in the measured adhesion energy – as determined from the pull-off force – between the two modes of approach. We conclude that the suppression of correlations does not result in the disappearance of the long-ranged attraction between the surfaces bearing randomly distributed charge domains. Correlation of opposing positive and negative charge domains on net-neutral surfaces has been both postulated(*7-11*) and calculated(*15*) to result in long-ranged attraction between them. But if, as we have shown, such correlation is not the reason, what then is the origin of this attraction, which is a hundred-fold larger than can be accounted for by dispersion forces alone?

We attribute the attraction to the asymmetry in the forces between equally- and between oppositely-charged surfaces across water (or other ion-containing liquids, see Methods). This is a counter-intuitive observation which has not earlier been remarked on. The interaction between two uniformly-charged surfaces is well described by the Poisson-Boltzmann (PB) equation, which for a 1:1 electrolyte (in the effectively 1D configuration of our experiment) is given by(*16*):

$$\frac{d^2\psi(x)}{dx^2} = \frac{2c_0 e}{\varepsilon \varepsilon_0} \cdot \sinh\left[\frac{e\psi(x)}{kT}\right] \quad (1)$$

where $\psi(x)$ is the potential at distance $x$ from the surface, $c_0$ is the concentration of ions in the bulk solution, $\varepsilon_0$ is the permittivity of free space, $\varepsilon$ is the dielectric constant, $k_B$ the Boltzmann constant, T the absolute temperature, and $e$ the electronic charge. This mean-field model, originally developed for the symmetric case of surfaces with a uniform, equal surface charge density, has been extensively validated experimentally, and has been extended in several directions(*4, 17-21*). The pressure between two approaching surfaces can be calculated by integration of the PB equation(*17*), yielding the net pressure $\Pi$ after the subtraction of the bulk osmotic pressure:

$$\Pi = c_0 kT\left(e^{\frac{-e\psi}{kT}} + e^{\frac{e\psi}{kT}} - 2\right) - \frac{\varepsilon \varepsilon_0}{2}\cdot\left(\frac{d\psi}{dx}\right)^2 \quad (2)$$

The first term corresponds to the net osmotic pressure in the gap (repulsive except when the potential vanishes) while the second term corresponds to the Maxwell stress which is always attractive. The equilibrium pressure between two similarly charged surfaces can be obtained by considering eq.(2) at the midplane of the intersurface gap where the gradient of potential is equal to zero, thus the only contribution to the pressure is the osmotic term. However, for two oppositely charged surfaces, the Maxwell stress never vanishes at finite separations. This result implies that two oppositely charged surfaces across water may attract more than two similarly charged ones will repel, for the same interacting areas, absolute surface charge densities and separations. This conclusion arises because of the qualitatively different mechanisms of repulsion and attraction across water between equally-charged and oppositely-charged surfaces, respectively. The former may be viewed as due to the osmotic pressure of trapped counterions at the midplane, while the latter may be seen as arising from the entropy gain upon release of counterion pairs from within the gap between the approaching surfaces(*16*).

To demonstrate that substantial long-ranged attraction between heterogeneous surfaces may be expected based on PB even without any correlation of opposing charges, we use a simple model, illustrated in the inset to fig. 4, based on the following assumptions: Randomly distributed charged domains on each surface; equal magnitude of charge density $|\sigma|$ on oppositely charged domains; a similar total area of the negatively and positively charged domains, for overall neutrality of each surface (a reasonable assumption, see fig. 1); and, in particular, domains that are substantially larger then the surface separation, enabling the neglect of domain edge effects and allowing us to treat the interactions as between flat uniform charge domains for which the PB equation applies. The dominant electrostatic interactions are then symmetric repulsive (+ vs. + and – vs. -, with resulting pressures $\Pi_{+/+}$), or antisymmetric attractive (+ vs. -; $\Pi_{+/-}$): It may readily be shown analytically (Methods) that, in the conditions of our experiments, $\Pi_{+/-} > \Pi_{+/+}$; the detailed behaviour requires numerical solution of the PB model.

This is shown in fig. 4, where the force vs. distance profile in our model (inset) is plotted from numerical solutions of the PB equation, under either constant charge or constant potential b.c.'s, incorporating also vdW attraction, using parameters typical for mica in water with no added salt, as for our experiments (figure 2). Also plotted (green band) is the spread of experimental profiles from fig. 2. We see clearly that our model – where there is no charge correlation and repulsive and attractive interactions cover equal areas - does indeed predict a net long-ranged attraction in the absence of any charge correlation (red and blue solid curves), though its range and magnitude are larger than our measured attractions (see below). We note that at small separations, it is constant potential rather than constant charge boundary conditions that apply (Methods), implying a monotonic attraction (red curve) – as observed experimentally – rather than the non-monotonic interaction (blue curve). Other calculated profiles (not shown) reveal, as expected, weaker net long-ranged attraction at higher salt concentration, and stronger attraction for larger $\sigma$.

Our PB-based treatment, while transparent and qualitatively demonstrating the long-ranged attraction, contains a number of assumptions: We ignore the effect of domain edges (which assumes all domains are much larger than the surface separation); and we assume that $\sigma_+ = |\sigma_-|$, and that the overall area of negative and positive charge patches is the same (fig. 1 indicates this is a reasonable assumption), so that each surface is overall neutral. However, the most significant factor responsible for the weaker measured attraction (green band in fig. 4) relative to the prediction of our model (red curve in fig. 4) is the fact that, as clearly seen in fig. 1, many of the charge patches are quite small, indeed smaller than the surface separation over much of the range D < 50 nm. Such patches, equivalent to quenched disorder of small charge patches, have been shown to result in little interaction at the PB level(*22*). Thus the effective area of the charge domains contributing to the attraction (those larger than ca. 200 nm say) is much smaller than the full coverage assumed in fig. 4 (solid red and blue curves), which explains qualitatively the much smaller measured attraction compared to the calculated one. When this is taken into account from the relative area of large domains, figure 1, we find a far closer fit to the data (broken red curve in fig 4). The essential point evident from fig. 4, however, is that there is no need for correlation of opposing charges to explain the strong attraction between randomly charged surfaces that we (and many others(*7-11*) earlier) have observed: it can be readily accounted for using the well-tested PB model.

Our findings may also account for the long-ranged attractions observed in very many studies between surfaces modified by physically-attached surfactant layers, which were previously attributed to 'hydrophobic interactions' (these experiments are reviewed in ref.(*23*)). It has long been pointed out(*24*) that such long-ranged attractions are strongly correlated with high mobility of the surfactant molecules: this in turn implies that the corresponding surfactant layers are able to rearrange in a facile manner. This correlation of long-ranged attraction with ease of surfactant-layer rearrangement makes it very conceivable that, under water, such layers undergo a thermodynamically-driven rearrangement to form random positive and negative charge patches similar to those observed in our study (fig. 1), and in earlier work. Taken together with our present findings that such random charge domains lead to long-ranged attraction even in the absence of any charge correlation (in contrast to earlier suggestions(*7-11, 15*) which all involved correlations), such a break-up would readily account for these long-ranged interactions earlier attributed to 'hydrophobic effects'. This is further supported by the observation – as seen in fig. 4 – that a rather small coverage of the surface by larger charge domains is sufficient to result in the observed long-ranged attractions. Moreover, the puzzling long-range attractions earlier observed between surfaces bearing nucleic acid residues(*25, 26*), attached to a physisorbed surfactant coating the substrate, may also be accounted for by a small extent of charge domain formation arising from the rearrangement of the underlying surfactant layer. In other words, our results suggest that it is not the 'hydrophobic effect' that is responsible for the long-ranged (ca. 20 nm – O(100 nm)) attraction noted between surfactant-coated surfaces, but, rather, a random, uncorrelated charge-patch formation arising from thermodynamically-driven surfactant rearrangements as discussed above.

In summary, we have found that long-ranged attractions between surfaces bearing random charge-domains across water, persist unchanged under conditions designed to frustrate correlated motion of charges as the surfaces approach. This shows directly that these attractions are not the result of correlations between oppositely charged domains, i.e. positive facing negative, developing on approach, as first proposed many years ago(*7-11, 15*). It is shown rather that such attractions may arise as the consequence of classically-understood electrostatic double-layer interactions (via the Poisson-Boltzmann equation) between large, random, uncorrelated charge domains on the opposing surfaces, once the counter-intuitive like-like vs. like-unlike interaction asymmetry is recognized. Our results immediately explain the recently-observed long-ranged attractions between surfaces made hydrophobic by surfactant monolayers that subsequently rearrange into positive and negative charged domains(*7-13*), with no need to invoke charge correlations. We believe they may account also for the many earlier long-ranged interactions measured between surfactant-coated surfaces and attributed to the 'hydrophobic interaction'(*23*). More generally, our findings shed new light on the stability of hydrocolloids and on interactions between heterogeneously-charged surfaces, which are ubiquitous in biological systems.

**Methods**:

*Surfactant coating*

Monolayers of the polar surfactant octadecyltrimethylammonium bromide ($CH_3(CH_2)_{17}N^+(CH_3)Br^-$, Sigma-Aldrich purity 98%+, used as received) were deposited on mica surfaces by self-assembly from aqueous solution at concentration 6-8-fold higher than the critical micelle concentration, at $70\pm5^0C$, well above the Krafft temperature for this system. The freshly cleaved molecularly-smooth mica sheets were dipped for 30 sec in the solution then rinsed in heated pure water. All water was purifed (Barnsted NanoPure, resistivity 18.2 $M\Omega.cm^{-1}$, nominal total organic content < 1 ppb). Layers were characterized by atomic force microscopy (AFM, Asylum MFP 3-D), contact angle goniometry (FTA200 goniometer) and surface force balance (SFB) measurements as described below. Following break-up of the monolayer on immersion in water, the overall area of the bilayers and their stability with time suggests little loss of the surfactant to solution.

*Surface Force Balance measurements*

The SFB enables the measurement of forces between two molecularly smooth mica surfaces (coated if appropriate, as in the present study) to be measured directly with angstrom-level resolution in the surface separation, and has been described in detail previously(*27*). In particular, it enables both normal motion between the surfaces and controlled lateral motion between them as the upper surface is moved parallel to the lower one via a sectored piezocrystal, as for the shear-approach mode described above. Force (F(D)) vs surface-separation (D) profiles were generated both manually, via a step-wise approach, and dynamically, where surfaces are made to approach at velocity $v_n$. The variation with time t of surface separation D(t) is determined via fast video recording, frame grabbing and analysis of the optical interference fringes of equal chromatic order. The corresponding force F(D) is then evaluated using the instantaneous balance of forces through the Newtonian relation(*28*):

$$F(D) = m\ddot{D} - K_n \delta D(t) + \frac{6\pi \eta R^2 \dot{D}}{D(t)}$$

where *m* is the total mass of the lower surfaces attached to the spring, $K_n$ is the spring constant: 200N/m, $\eta$ is the dynamic viscosity of water, *R* is the mean radius of curvature of the surfaces. The time derivative of the data, $\dot{D}$, at large separation is constant and equal to the approach velocity, $v_n$ of the surface due to the applied approach motion (ranging from 4 - 20 nm/sec in different runs). Any deviation from this slope is due to the deflection $\delta D(t)$ of the spring connected to the lower lens $\delta D(t) = D_{t=0} - D(t) + v_n t$. The last term in the equation for *F(D)* is the hydrodynamic force given by the Taylor equation under the assumption of no slip boundary condition.

*Electrostatic double-layer interactions*

Analytical solutions of eq (2) for equally and for oppositely charged surfaces may be obtained for certain regimes determined by the relation between the length scales in the PB model(*17, 20*), the so-called Gouy-Chapman length *b* and the Debye screening length $\lambda_D$. These are given by $b = \frac{1}{2\pi l_B \sigma}$ and $\lambda_D = \sqrt{\frac{\varepsilon\varepsilon_0 k_B T}{2 c_0 e^2}}$, where $l_b$ is the Bjerrum length, defined as the separation between elementary charges at which the electrostatic interaction equal to the thermal energy: $l_b = \frac{e^2}{4\pi\varepsilon\varepsilon_0 kT}$ (= 7Å for water at 300K), which yields $b$ = 160Å (evaluated at $\sigma = (e/70\text{nm}^2)$, a typical value for mica). $\lambda_D$ in purified water with no added salt can range from about 50nm up to about 1μm (depending on the amount of dissolved carbon dioxide and ions leaching from the glassware in the water). The Gouy-Chapman regime(*17, 20*), which is when $\lambda_D >> D$ and $b << D$, is most relevant for our experiments in water with no added salt where the long-ranged attractions are observed experimentally. The expressions for the pressure between the surfaces in this regime are(*17*) $\Pi_{+/+} \cong \frac{\pi}{2 l_B D^2}$ and(*20*) $\Pi_{+/-} \cong -\frac{2}{\pi l_B D^2} \ln^2\left(\frac{D}{8\lambda_D}\right)$, giving the ratio *Ra* between the repulsive and attractive interactions in this regime as $\left|\frac{\Pi_{+/-}}{\Pi_{+/+}}\right| = Ra \cong \frac{4}{\pi^2}\ln^2\left(\frac{D}{8\lambda_D}\right)$.

*Ra* > 1 throughout the Gouy-Chapman range, increasing significantly for D < $\lambda_D$/2. This implies that in this regime – corresponding to our experiments in no added salt water - the attraction will dominate the repulsion (the terms defining *Ra* were derived under constant charge boundary conditions (b.c), but – as seen in the numerical simulations – the ratio of attractive to repulsive pressure is even higher when constant potential b.c. are used). On physical grounds, one expects constant charge b.c. to apply better at large D, and constant potential b.c. to apply more closely at small separations where the forces are larger and charge regulation is more likely; see also fits in fig. 2. (While in this paper we focused on double layer interactions between charged surfaces across aqueous media, the most common case, the asymmetry between the magnitudes of attraction and repulsion is expected also across other ion-containing liquids, provided that double layer interaction is taking place(*29*))

**Acknowledgments**

We thank Sam Safran for useful discussions. Support from the Israel Science Foundation, from the European Research Council and from the Minerva Foundation is acknowledged with thanks.

**Author contributions**

GS designed, performed and analyzed experiments, developed the interaction model and wrote the paper; D.B-Y. wrote the algorithm to generate profiles from the PB equation; YD performed AFM imaging; SP and NK performed experiments; JK initiated and directed the project and wrote the paper.

**Figure Captions**

Figure 1: Surfaces with random charge domains. AFM tapping-mode images of the model surfactant (octadecyltrimethylammonium bromide) on mica: A: in air, B: after 22 hrs immersion time in pure water. The cartoons illustrate the break-up of the initially uniform monolayer into positively-charged bilayer domains on a negatively-charged substrate. For the micrograph shown the bilayer domains (white, positively charged) cover some 45% of the total area in B, while those over lateral dimensions > 200 nm cover 7% of the total area.

Figure 2: Normalised force vs. distance profiles ($F_n(D)/R$ between surfaces with random charge domains, in the Derjaguin approximation, where R is the surface radius of curvature) between random charge-domain-coated surfaces (arising as in fig. 1B) on mica, carried out both in the straight-approach mode (blue profiles) and in the shear-approach mode (red profiles). Surfaces were immersed for 40 and 43 hours in water, and force profiles from two independent contact points are shown (typical of many others); the upper insets show the profiles on a magnified scale, where the broken curves are the van der Waals interaction alone (Hamaker constant as below). Consecutive force profiles alternating between shear-approach (red) and straight-approach (blue) were carried out at each of the contact points. Amplitudes of shearing during 'shear approach' were ca. 600nm or more with frequencies up to 4Hz resulting in shear velocities of 3.8-5.1μm/sec. First approach in both contact points was carried out using 'straight approach' (inset A), followed by a 'shear approach' (inset B). A control profile between two bare mica surfaces (prior to coating with STAB) across water is shown (■) and fitted to the solution of equation (2) together with the vdW interaction (with Hamaker constant $A=2 \times 10^{-20}$ J ) under limits of constant potential

and constant charge boundary conditions (lower and upper grey curves respectively). The corresponding parameters are: surface potential $\psi_0$= -117mV, Debye length $\lambda_D$=75nm, surface charge density $\sigma = e/70\text{nm}^2$. The data points, intermediate between the two curves, suggest boundary conditions somewhere between these limits.

Figure 3: Surface energy of the random charge-domain-coated surfaces (as in fig. 1). Surface energies are shown as function of time in water and of multiple approaches at given contact point, for several different contact points and two independent experiments, calculated via the force to separate the two adhered surfaces using the Johnson-Kendal-Roberts model(*30*). Consecutive approaches at a given contact point are alternately in the straight- and in the shear-approach modes. (●)- first approach at a contact point. (■)- 2$^{nd}$ entry at a contact point. (▲)-3$^{rd}$ entry at a contact point. (▼)- 4$^{th}$ entry at a contact point.

Figure 4: Numerical solutions for the interaction between two surfaces with uniform and with random charge domains. Results are based on the PB equation (1) combined with vdW attraction, under constant charge (broken blue curves) and under constant potential (broken red curves) boundary conditions. The upper (repulsive) broken curves show the interaction between two equally charged surfaces (symmetric interaction). The lower (attractive) broken curves show the interaction between two oppositely charged surfaces (antisymmetric interaction). The thicker smooth curves represent half the sum of both interactions (i.e. symmetric interaction/2 + antisymmetric interaction/2), which corresponds to that expected from our model of random charge domains described in the text. The green shaded region corresponds to the range of experimental attractions in fig. 2. The inset cartoon illustrates the model,

where outward pointing arrows indicate like-like repulsion, and inward pointing ones are like-unlike attraction. The calculations were carried out for surface charge density of $|\sigma_+|=|\sigma_-|=1/70\text{nm}^2$ (constant charge) or for $\psi_0 = -117$ mV (constant potential), with Debye length $\lambda_D = 75$ nm and Hamaker constant $A=2\times10^{-20}$J for the van der Waals attraction. The thick broken red curve corresponds to the net interaction, where only 5% of the domains contribute to the interaction, as suggested from the relative areas of the large domains in fig. 1 (calculated with constant potential b.c.'s, see text).

Fig1

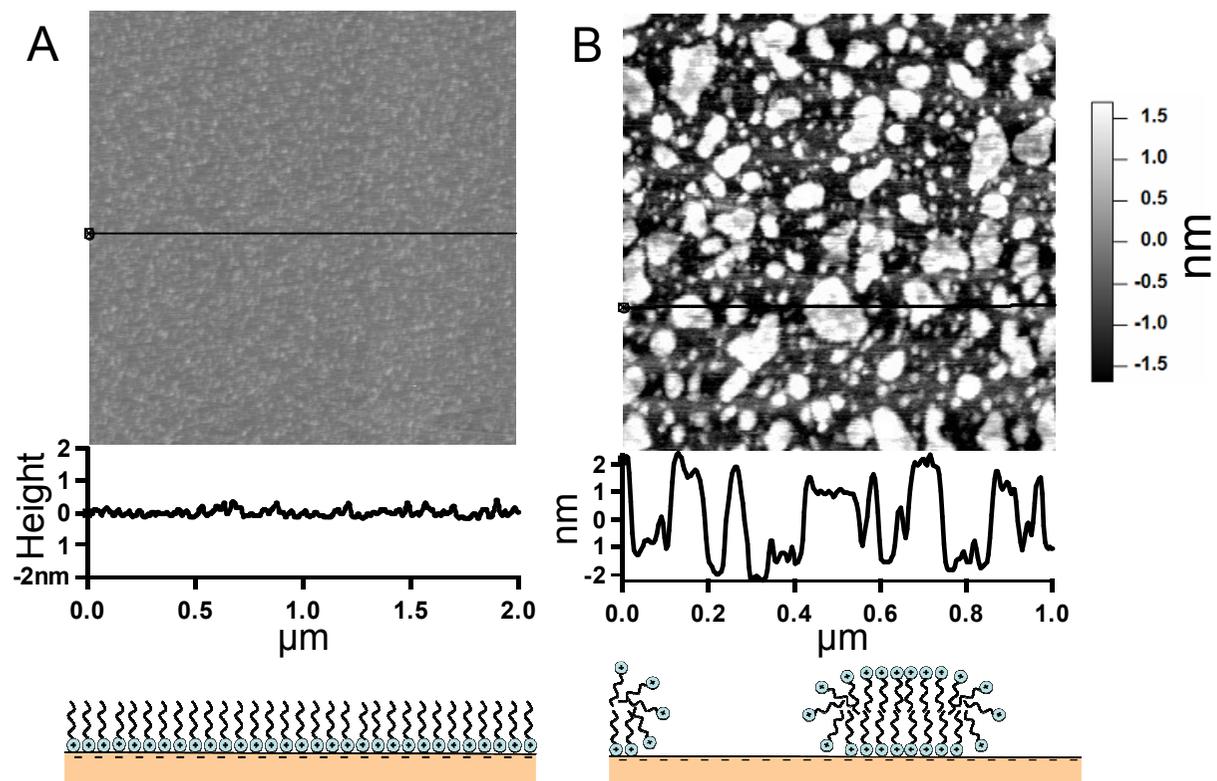

Fig 2

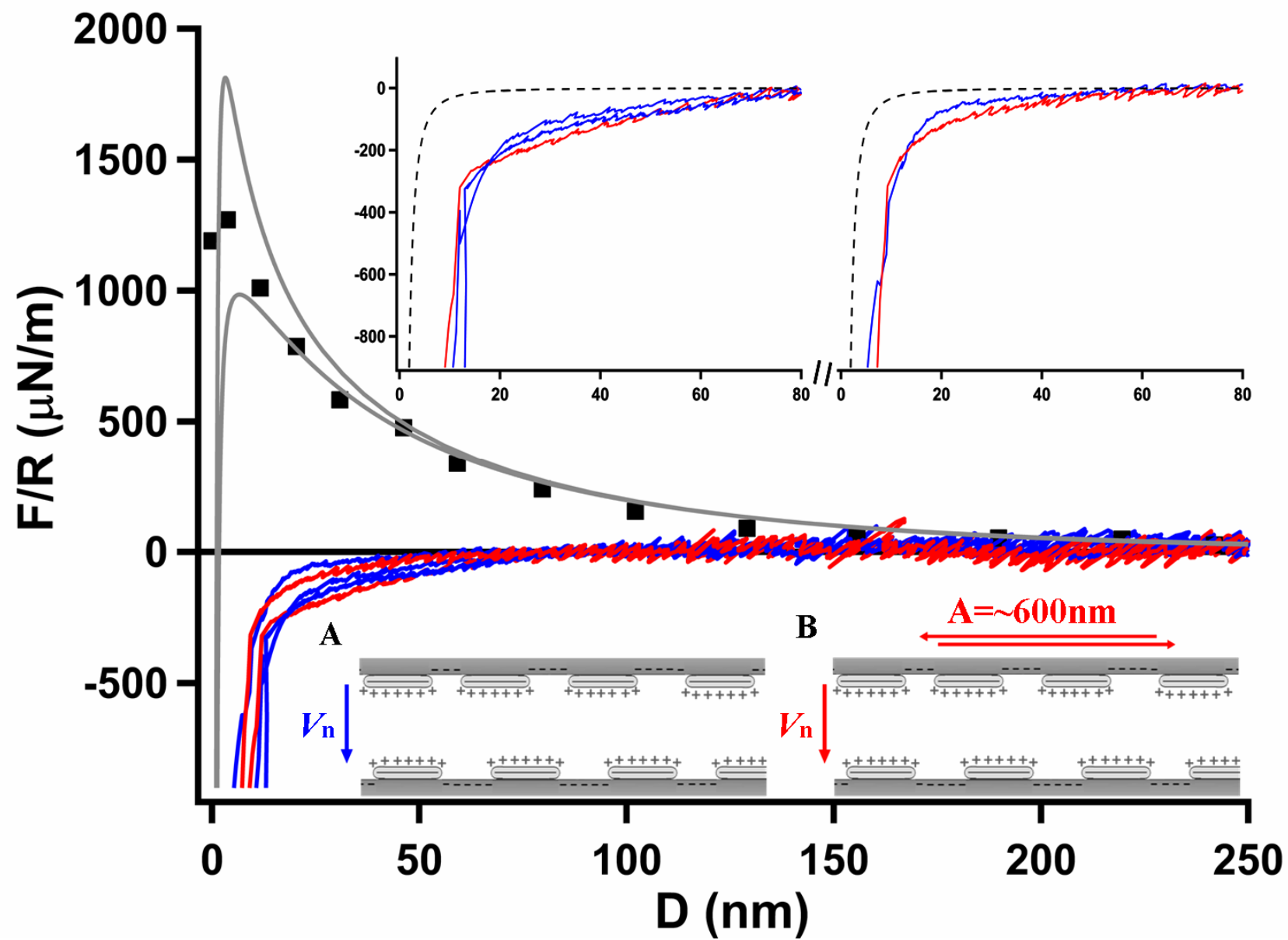

Fig 3

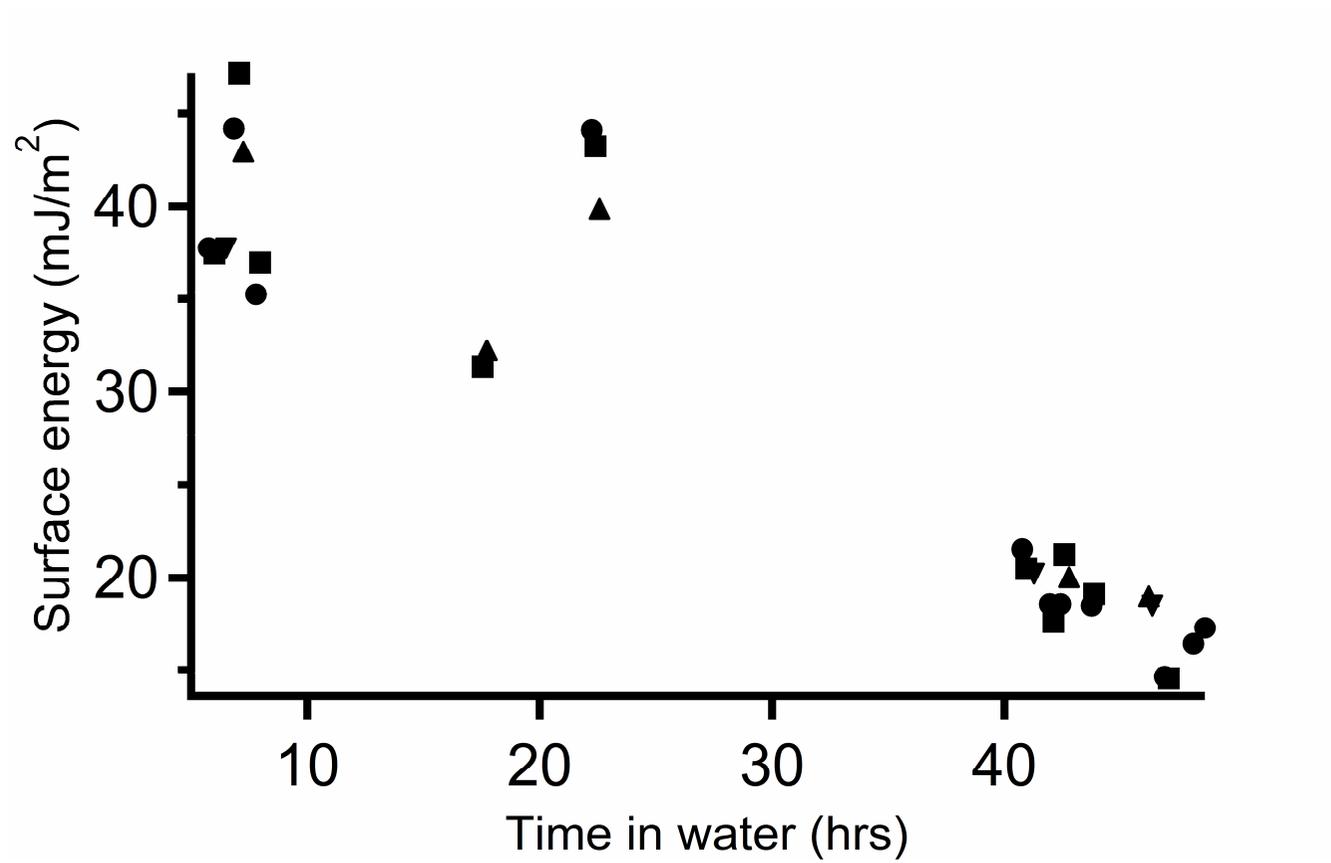

Fig 4

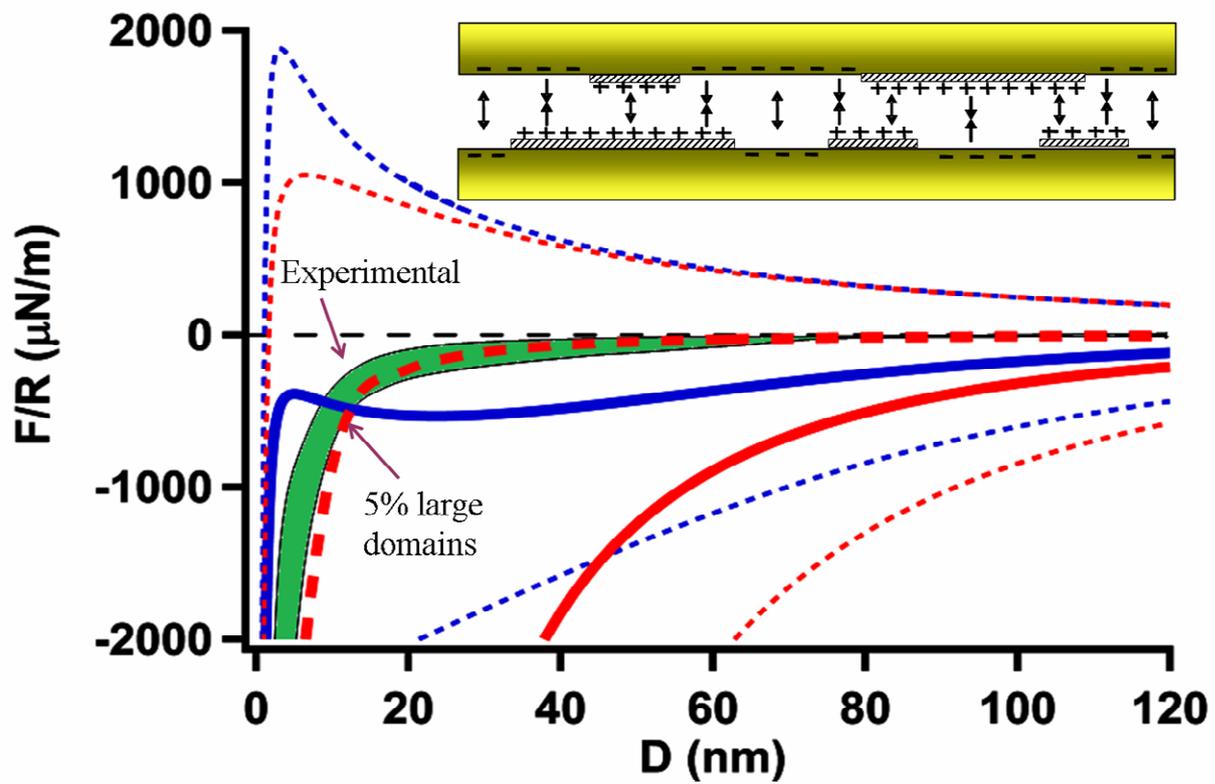